\documentclass{iaus}
\usepackage{graphicx}
\usepackage{epsfig}

\title[$s$-elements in PNe] 
{Abundances of $s$-process elements in planetary nebulae: Br, Kr \& Xe}

\author[Zhang et al. ]   
{Y. Zhang$^{1,2}$, R. Williams$^2$,
E. Pellegrini$^{3}$, K. Cavagnolo$^{3}$, J. A. Baldwin$^{3}$,
B. Sharpee$^{4}$, M. Phillips$^{5}$ \and X.-W. Liu$^{1}$}

\affiliation{$^1$Department of Astronomy, Peking University,
Beijing 100871, P. R. China\\ $^2$Space Telescope Science Institute, 3700 San Martin Drive, Baltimore, MD 21218, USA\\ $^3$Department of Physics \& Astronomy,
Michigan State University, East Lansing, MI 48824, USA\\ $^4$SRI International, 333 Ravenswood Ave.
Menlo Park, CA 94025, USA\\ $^5$Las Campanas Observatory, Carnegie Observatories,
Casilla 601, La Serena, Chile}

\pubyear{2006}
\volume{234}  
\pagerange{111--111}
\date{?? and in revised form ??}
\jname{Planetary Nebulae in our Galaxy and Beyond}
\editors{M. J. Barlow \& R. H. M\'endez, eds.}
\begin{document}

\maketitle

\begin{abstract}
We identify emission lines of post-iron peak elements in very high signal-to-noise spectra of a
sample of planetary nebulae.  Analysis of lines from ions of Kr and Xe
reveals enhancements in most of the PNe, in agreement with the theories
of $s$-process in AGB star.
 Surprisingly, we did not detect lines from Br even though $s$-process calculations
indicate that it should be produced with Kr at detectable levels.

\end{abstract}

\firstsection 
\section{Introduction}

As remnants of AGB stars, planetary nebulae (PNe) represent material that
has undergone nuclear processing in their precursors via the $s$-process.
The analysis of nebular emission provides essential information for stellar
models. However, detection of emission lines from $s$-elements has been
hampered by their weakness and by uncertainties in the atomic data.
A pioneer attack on the problem was made by Pequignot \& Baluteau (1994),
who identified a number of post-Fe emission lines in NGC\,7027.
Dinerstein and collaborators studied $s$-process elements in PNe by
searching for their IR emission (Dinerstein 2001; Sterling \& Dinerstein 2004,
this volume) and far-UV absorption (Sterling et al. 2002, 2003). As part
of an on-going program to detect weak lines in PNe, we have obtained
a number of high resolution spectra of very high S/N's.

\section{Observations and Analysis}

Our current sample of objects consists of four PNe (IC~2501, IC~4191,
NGC~2440, and NGC~7027). The spectra were obtained with the KPNO
4m using the Cassegrain echelle spectrograph and the LCO 6.5m with
the MIKE echelle spectrograph at a resolving power of about 25,000.
Fig.~1 shows the quality of our spectra. The deep spectra enable us
to detect extremely faint lines with a flux of  
$\sim10^{-6}$ the intensity of H$\beta$. For comparison, we also
searched for Kr and Xe in two H~{\sc ii} regions, the Orion Nebula 
(Baldwin et al. 2000) and NGC~3576 (Garc{\' i}a-Rojas et al. 2004).

We have detected krypton and xenon emission lines, [Kr~{\sc iii}] $\lambda6827$,
[Kr~{\sc iv}] $\lambda\lambda5868,5346$, [Kr~{\sc v}] $\lambda6256$,
[Xe~{\sc iii}] $\lambda5846$,
[Xe~{\sc iv}] $\lambda\lambda5709,7535$, and [Xe~{\sc v}] $\lambda7077$.
However, we failed to detect Br lines, [Br~{\sc iii}] $\lambda6133$ and
[Br~{\sc iv}] $\lambda7368$, even though the current $s$-process
calculations indicate that Br should be produced along with Kr at
detectable levels. The reason remains unknown.

Using the current available atomic data, we have determined Kr and Xe 
abudances. We converted ionic ratios to elemental values 
 by making use
of the similarity in ionization potentials of the noble gases, from which it
follows that
${\rm (Kr, Xe)}/{\rm Ar}=[({\rm Kr}^{+2}, {\rm Xe}^{+2})+({\rm Kr}^{+3},
{\rm Xe}^{+3})]/({\rm Ar}^{+2}+{\rm Ar}^{+3})$ and
$[({\rm Kr}^{+2}, {\rm Xe}^{+2})+({\rm Kr}^{+3}, {\rm Xe}^{+3})+({\rm Kr}^{+4},
{\rm Xe}^{+4})]/ \\
({\rm Ar}^{+2}+{\rm Ar}^{+3} +{\rm Ar}^{+4})$
for low- and high-excitation PNe, respectively.
The upper limits for the Br abundances were estimated.
These results, relative to the solar values, are given in Table~1.
It is evident that Kr and Xe are both
enhanced by similar factors of up to 10 in the five PNe, but not in the two H~{\sc ii}
regions which represent unprocessed ISM gas.

\begin{figure}
\begin{center}
\epsfig{file=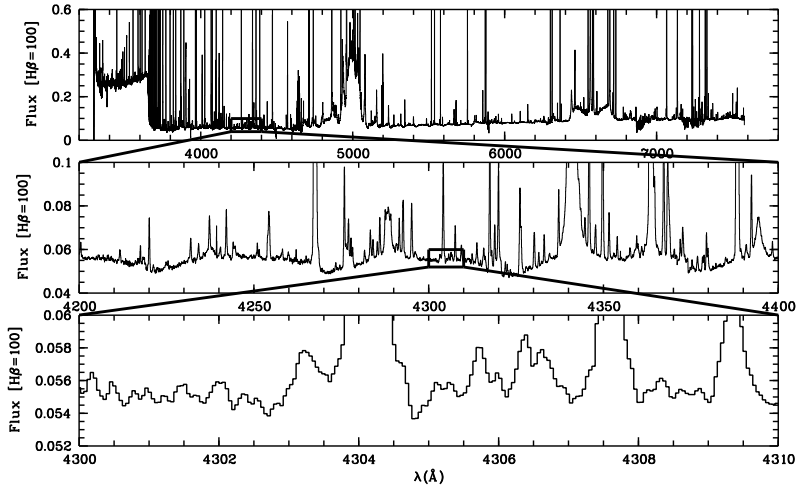,
height=6.5cm, bbllx=41, bblly=313, bburx=558, bbury=630, clip=, angle=0}
\caption{Spectrum of IC~2501 illustrating the quility of our spectra.}
\end{center}
\end{figure}

\begin{table}
 \begin{center}
\caption{Abundances of Br, Kr, and Xe.}
\begin{tabular}{lllllllll} \hline
            & \multicolumn{5}{c}{PNe} & & \multicolumn{2}{c}{H~{\sc ii} regions} \\
\cline{2-6}  \cline{8-9}\\
{Abundance} &{IC~418} & {IC~2501} & {IC~4191} & {NGC~2440} &
{NGC~7027} & & {Orion} & {NGC~3576}  \\
\hline
\vspace{0.0001in}[Br/Ar] & $<-0.8$&  $<-1.2$&  $<-0.5$&  $<-1.0$& $<0.3$ & & ... & ...\\
\vspace{0.0001in}[Kr/Ar] & 0.76   & 0.04    & 0.38    & 0.15 & 1.04 & & $-0.08$ & $<-0.69$ \\
\vspace{0.0001in}[Xe/Ar] & 0.91   & 0.01    & 0.51    & 0.33 & 0.87 & & $<0.58$ & ... \\
\hline\end{tabular}
\end{center}
\end{table}

\end{document}